\documentclass[12pt]{article}

\usepackage{epsfig}

\newcommand{\eq}{\begin{equation}}
\newcommand{\eqx}{\end{equation}}
\newcommand{\eqn}{\begin{eqnarray}}
\newcommand{\eqnx}{\end{eqnarray}}
\newcommand{\ad}{a^{\dagger}}

\begin{document}

\title{\Large\bf
\begin{flushright}
\normalsize\rm 4/TPJU-7/03
\end{flushright}
\vskip 20 pt
{\Large Quantum systems in a cut Fock space}  }

\author{\large M. Trzetrzelewski  and J. Wosiek\\  \\  
  $M.\; Smoluchowski\; Institute\; of\; Physics$\\
$Jagellonian\; University $\\ 
$Reymonta\; 4\;, 30-059\; Krakow,\; Poland\;$\\
}

\maketitle

\begin{abstract}
Standard quantum mechanics is viewed as a limit of a cut system with artificially restricted dimension 
of a Hilbert space. Exact spectrum of cut momentum and coordinate operators is derived and the limiting
transition to the infinite dimensional Hilbert space is studied in detail. The difference between systems with 
discrete and continuous energy spectra is emphasized. In particular a new scaling law, characteristic
for nonlocalized, states is found. Some applications for supersymmetric quantum mechanics are briefly outlined.
\end{abstract}
\newpage

\section{Introduction}
Recently there has been a further progress in understanding a family
of quantum mechanical systems emerging from the dimensional reduction
of supersymmetric gauge theories first studied in \cite{CH,HS}. 
Resulting models, even though much simpler than the original
field theories, can be rather complex and with nontrivial solution. 
In the programme proposed in \cite{JW}, the hamiltonian
is diagonalized in the limited Hilbert space providing the complete spectrum
of the system with a cutoff. Then the cutoff is gradually removed and convergence
towards the exact (i.e. the infinite Hilbert space) spectrum is observed.
Interestingly, even for quite complex systems with 15 degrees of freedom the convergence
occurs before the size of the basis grows out of control \cite{JW2,KW,CW}.

     Above method, which is essentially numerical, raises a number of  
theoretical questions about exact solutions of quantum mechanics in a limited Hilbert space.
Some of them will be addressed in this paper. In the next Section we define the cutoff as 
the maximal number of harmonic oscillator quanta, $N$. Then we solve analytically for the 
spectrum of the momentum and coordinate operators at arbitrary finite $N$. In Section 4 
the asymptotic behaviour for large $N$ is derived and the new scaling law required 
to recover the infinite Hilbert space limit is formulated. 
The universality of the above scaling and similarity of the whole procedure to the familiar 
continuum limit in lattice field theories is discussed. 
It is also shown
that present results prove the assertion of Ref.\cite{JW} that 
the continuous spectrum in quantum mechanics gives rise to the power-like dependence on the cutoff.
On the other hand, it was found numerically, that the eigenenergies of the discrete, localized states have much faster convergence.
This correlation between the nature of the spectrum and the cutoff dependence is very useful in studying
 supersymmetric systems where the distinction between
continuous and discrete spectra is an important issue \cite{LNDW,BFSS}. 

A general idea of increasing the number of states in a trial wave function is of course
not new and is closely connected with the classical variational calculations\footnote{For some applications
for gauge systems see e.g. \cite{L,LM,VABA,vBN}.}.
However we are not aware of any systematic studies of the cutoff dependence based on the exact
solutions of cut systems, especially in the context of contrasting the discrete and continuous spectra.
 
\section{A cut Fock space}

Every quantum hamiltonian can be written in terms of the creation and annihilation
operators of a simple, normalized harmonic oscillator
\eq
H(Q,P)=H(a,\ad),  \label{ham}
\eqx
where
\eq
Q=(a+\ad)/\sqrt{2},\;\; P=(a-\ad)/i\sqrt{2},\;\;\;[a,\ad]=1.
\eqx
And similarly for the fermionic degrees of freedom if required. It follows that a large class of
polynomial hamiltonians has a simple representation in the eigenbasis $\{|n>\}$
of the occupation number operator $\ad a$.
\eq
|n>={{\ad}^n\over \sqrt{n!}} |0>.   \label{bas}
\eqx
Apart from  the theoretical simplicity, this basis has an important practical advantage.
It is labeled by a discrete index $n$ and consequently is well suited for numerical
applications. 

     Therefore a straightforward method for numerical solution of any quantum problem consists of: 
a) limiting the basis, Eq. (\ref{bas}), to e.g. $n \le N$; b) calculating the finite matrix representation of the hamiltonian
(\ref{ham}) in such a cut basis; and c) numerically diagonalizing above finite matrix to obtain complete spectrum
and the eigenstates of the system. Moreover, changing the cutoff $N$ provides the model independent way to estimate
the systematic errors introduced by limiting the number of allowed quanta.

     It is evident that theoretical understanding of the cutoff dependence for various quantum mechanical systems
would further extend applicability of this approach. In particular, the knowledge of the asymptotic behaviour
of the eigenvalues of a cut hamiltonian for large $N$ could be used to perform quantitative extrapolation to 
the infinite cutoff limit. 
We begin studying these questions with the simplest building blocks of any hamiltonian, i.e. the $P$ and $Q$ operators.
\section{The spectrum of cut momentum and coordinate operators}
     
Standard expressions for the matrix elements of the $P$ and $Q$ operators in the occupation 
number basis read. 
\eqn
\langle n \mid P \mid  k \rangle &=& 
\frac{1}{i}\sqrt{\frac{k}{2}}\delta_{n,k-1}-\frac{1}{i}\sqrt{\frac{k+1}{2}}\delta_{n,k+1},\\
     \langle n \mid Q \mid  k \rangle  
  &=& \sqrt{\frac{k}{2}}\delta_{n,k-1}+\sqrt{\frac{k+1}{2}}\delta_{n,k+1}.  
\eqnx
 In the Hilbert space, limited to maximum $N$ quanta, $P$ and $Q$ become $(N+1)\times(N+1)$ 
matrices. Hence, the eigenvalues, $\lambda$, of e.g.  momentum, are given by the zeros of 
the following determinant

\begin{equation} 
 I_{N+1} = \left\vert\begin{array}{cccccc}
        -\eta\    & \sqrt{1}  & 0        & .         & .        & .            \\
        -\sqrt{1} & -\eta     & \sqrt{2} & .         & .        & .             \\
        0         & -\sqrt{2} & -\eta    & .         & .        & .             \\
        .         & .         & .        & .         & .        & .            \\
        .         & .         & .        & .      & -\eta       & \sqrt{N}      \\
        .         & .         & .        & .      & -\sqrt{N} &-\eta    
          
             \end{array}\right\vert ,
\end{equation}
where $\eta=i\sqrt{2}\lambda$.  
Laplace expansion gives, for the simpler variable
$J_n=I_n/n!$, the following recursion relation
\begin{equation}  
(n+2)J_{n+2} + \eta J_{n+1} - J_{n} = 0,\ \ J_{1}=-\eta,\ \ J_{2}=\frac{1}{2} ({\eta}^{2} + 1),  
\end{equation}
which is closely related to the well known recursion for the Hermite polynomials. We find
\eq
I_n(\lambda)=2^{-{n\over2}} i^n H_n(-\lambda).
\eqx
It follows that the spectrum of the momentum in the cut Hilbert space with maximum $N$ quanta
 is given by the 
zeroes of the Hermite polynomials $H_{N+1}(z)$
\eq
  p^{N}_{m}=z^{(N+1)}_{m}, \ \ \ \   \hbox{where} \ \ \ \   H_{N+1}(z^{(N+1)}_{m})=0 ,
 \ \ \ \ m=1,2,\ldots,N+1. \label{spec} 
\eqx 
This constitutes the main result of present paper.

Calculations for the coordinate operator $Q$ are very similar.
Resulting eigenvalues, $q^{N}_{m}$, are identical 
\eq
  q^{N}_{m}=z^{(N+1)}_{m},
\eqx  
which could have been expected from the duality between $P$ and $Q$. 
It is important that this duality
is not violated by our cutoff. Of course the eigenvectors of $P$ and $Q$ are different.
Since the eigenvalues are symmetric with respect to the origin, we will use only positive
ones. Hence, we introduce a slightly different enumeration 
\eqn 
 p^{N}_{0} < p^{N}_{1}<p^{N}_{2}<\ldots < p^{N}_{m} < \ldots \le p^{N}_{N/2},  & N - even,   \label{pos2} \\
 p^{N}_{1} < p^{N}_{2}<\ldots < p^{N}_{m} < \ldots \le p^{N}_{(N+1)/2}, &  N - odd. \label{pos1}  
\eqnx  
which will be used in the next Section.

\section{Continuum limit - scaling}

Restricting the infinite Hilbert space by artificially
cutting off the basis of the Fock states has some resemblance to the familiar
discretization in a lattice field theory. In both situations a cutoff makes
a problem numerically tractable and in the end it has to be removed to reach
a physical description of the continuous world. In lattice 
calculations this is much more delicate due to the infinite number of degrees of freedom,
and must be accompanied with a nontrivial adjustment (scaling) of bare parameters.
Interestingly, similar phenomenon is observed when we increase the size of the Hilbert space
in our cut quantum mechanics in order to recover standard quantum system. To see this, consider
the large N limit of the spectrum of the momentum operator derived in the previous Section.
From the known asymptotics of the roots of Hermite polynomials \cite{ABR} we obtain 
(see Appendix)
\eqn
p_{m}^{N}&=&\frac{\pi m}{\sqrt{2N+3}} \sqrt{ 1+\frac{{{\pi}^{2}}m^{2} -3/2}{3(2N+3)^{2}} }
+ O(N^{-4.5}), \label{eva1} \\  
 p_{m}^{N} &=&  \frac{\pi(m-\frac{1}{2})}{\sqrt{2N+3}} 
\sqrt{ 1+\frac{{{\pi}^{2}}(m-\frac{1}{2})^{2} - 3/2}{3(2N+3)^{2}} }+O(N^{-4.5}), \label{evals}  
\eqnx
with $m$ and $N$ as in (\ref{pos2}) and (\ref{pos1}) respectively.
Therefore an infinite N limit {\em at fixed m} is trivial and does not reproduce 
the spectrum of the P operator in the infinite Hilbert space.
 On the other hand varying the "principal quantum number" $m$ with the cutoff $N$
according to the following prescription
\eq
m={\sqrt{2N}\over \pi}p, \label{scal} 
\eqx
defines the scaling limit of our quantum system 
\eq
lim_{N \to \infty} p_{m(N,p)}^{N}=p, \label{conti}
\eqx
which reproduces the known spectrum of the momentum for any real eigenvalue $p$. 
Because of the above analogy with the lattice approach,
the adjustment (\ref{scal}) will be called scaling and we will refer to the scaling limit, 
Eq.(\ref{conti}), as the continuum limit. 

A number of comments is in order here. First, for finite cutoff $N$, the $N+1$ eigenvalues, 
Eq.(\ref{evals}), span the interval
of the length $O(\sqrt{N})$ and are separated by the distance $O(1/\sqrt{N})$. 
When $N$ tends to infinity
they spread out covering the whole real axis and, at the same time, become more dense
approximating any real number $p$ with arbitrary precision. In another words, limiting the 
number of quanta
can be thought of as regularizing the system both in the infrared and in the ultraviolet.
This means that the single dimensionless parameter $N$ controls both the continuum limit and 
the infinite volume limit at the same time.
   
    Second, the scaling, Eq.(\ref{scal}), is universal, i.e. it applies 
to the spectrum of any operator which
commutes with $P$. For example, it was found \cite{TRZ} that the spectrum of a cut free particle hamiltonian, 
$H=P^2/2$, behaves at large N as
\begin{eqnarray} 
E_m^N&=&\frac{{\pi}^2}{2} \frac{(m-1/2)^2}{2N+5}, \nonumber \\ 
E_m^N&=&\frac{{\pi}^2}{2} \frac{m^2}{2N+5}, 
\label{eigenen}
\end{eqnarray}
for $N$ even and odd respectively. This implies, together with (\ref{scal}),  the standard spectral relation 
$E(p)=p^2/2$ in the continuum limit. 
Note that the scaling deduced only from the momentum variable was sufficient to derive the continuum 
limit of the energy
\footnote{Of course we also needed the solution of the free particle problem in a cut Fock space.}.
 
Third, this universality should also extend to any scattering problem in quantum mechanics, 
that is to less trivial hamiltonians with continuous spectra, provided the momenta can be defined 
asymptotically.
This observation is particularly useful for studying nonlocalized solutions of supersymmetric systems.

Finally we stress the difference in the cutoff dependence between the eigenenergies of 
the localized and nonlocalized states.
Results (\ref{evals}) and (\ref{eigenen}) prove that for the continuous spectrum the eigenenergies
fall as a power of the cutoff.
On the other hand there is a strong numerical evidence that in the case of the discrete, 
localized eigenstates the convergence
of the eigenenergies is much faster, probably exponential or even stronger . 
The difference between the two
is so dramatic that it was successfully used to distinguish between both classes of solutions 
from the finite N data alone \cite{JW,JW2,KW,CW}.
The scaling (\ref{scal}) applies only to the continuous spectrum. 
Continuum limit of the eigenenergies from the discrete spectrum
is achieved simply by taking large $N$ at fixed $m$. 

To conclude this Section we derive the form of the large $N$ corrections to the 
continuous spectrum. Let us use a slightly more general form of (\ref{scal})
\eq
m={\sqrt{2N}\over \pi}p+b, \label{scal2} 
\eqx
with arbitrary real $b$.
Inserting Eq.(\ref{scal2}) into (\ref{eva1}) and regrouping terms according 
to their N dependence gives
\begin{eqnarray} 
p_{m}^{N} = p &+& \pi(b-\frac{1}{2})\frac{1}{\sqrt{2N}} + 
\frac{1}{6} p(p^2 -9)\frac{1}{2N} \\
 &&+ \frac{\pi}{2}(b-\frac{1}{2})(\frac{2}{3}p^2 -3)   \frac{1}{(2N)^{\frac{3}{2}}}+\ldots  
\end{eqnarray}
The parameter $b$ can be used to speed up the convergence. 
Obviously $b=1/2$ is optimal in this case, however this choice 
is not universal in general. In fact we readily see this even in our example.
For even number of quanta, Eq.(\ref{eva1}), the best choice is $b=0$. 
The same remark applies if one improves the convergence by replacing
$\sqrt{2N}$ by , e.g. $\sqrt{2N+3}$ in (\ref{scal}). Of course the limiting value
of the momentum is independent of $b$ in agreement with the universality.

\section{Summary and outlook}

Quantum mechanics in a cut Hilbert space is an interesting theoretical framework
admitting a number of exact solutions which, to our knowledge, have never been
obtained before. Apart from its theoretical and pedagogical significance, 
analytical understanding of the cutoff dependence of the solutions provides the useful
practical tool to study more complex systems.  

Scaling found in the spectrum of the 
cut momentum operator seems to be a general property of nonlocalized states.
It was confirmed in the simple case of a free particle in one dimension and we are currently
extending this to higher dimensions. 
Many other applications, with various degree of complexity, are possible. 
One of the most interesting is the study
of the scattering states in the family of supersymmetric Yang-Mills quantum mechanical
 systems with the flat valleys potentials (SYMQM).
Reduced from the two dimensional space-time (D=2) theory, SYMQM is essentially 
the quantum mechanics of a free
particle in three (color) dimensions with a constraint \cite{CW}. Thus it can be readily analyzed with 
the present method.
The D=4 system is already nontrivial, revealing both the discrete and continuum spectrum \cite{JW}. 
Since one can assign a momentum along the valley to the asymptotic scattering
sates, we expect that above scaling should also work in this case. 
Recent numerical progress in solving this system in a cut Hilbert space makes this 
programme rather feasible now.
Finally, in the most complex D=10 SYMQM, with its threshold bound state and the continuum
of the scattering states, one can also assign an asymptotic momentum to these states \cite{BERS,PW}.
Hence the momentum scaling found here should be used in extracting
the continuum limit of this model from its cut formulation.
\vspace*{.3cm}
 
\noindent {\em Acknowledgments} 
JW thanks the Theory Group of Pisa University for their hospitality. 
This work is  supported by the
Polish Committee for Scientific Research under the grant no. PB 2P03B09622,
during 2002 -2004.

\section{Appendix}

Here we will derive the asymptotic form of the zeroes of the Hermite polynomials $H_n$ for
large order $n$. When $n$ is even they can be obtained from the following relation  \cite{ABR}
\begin{equation} 
H_{n} (z) = (-1)^{ \frac{n}{2} } 2^{n} { \frac{1}{2} n }! L_{{ \frac{n}{2} }}^{-\frac{1}{2}} (z^{2}),   
\end{equation} 
 where $L_{n}^{\alpha} (z^{2})$ are the generalized Laguerre polynomials with parameter $\alpha$. 
Then, roots of the Laguerre polynomials are approximated by the first $n/2$ roots of the Bessel functions.
Let 
 $z_{m}^{n}$, $t_{m,\alpha}^{\frac{n}{2}}$ and $j_{m,\alpha}$ denote the m-th positive root of the
$H_n(z)$, ${L_{\frac{n}{2}}}^{\alpha}(z)$ and the Bessel function $J_{\alpha}(z)$ respectively.
One has \cite{ABR}
\begin{equation}  
t_{m,\alpha}^{\frac{n}{2}}= 
\frac{{j_{m,\alpha}}^{2}}{4k_{\frac{n}{2},\alpha}} 
\left( 1+\frac{2({\alpha}^{2} - 1)+{j_{m,\alpha}}^{2}}{ 48{k_{\frac{n}{2},\alpha}}^{2}} \right)+O(n^{-5})  ,
\end{equation}
 where 
\begin{equation} 
   k_{\frac{n}{2},\alpha}=\frac{n}{2}+\frac{\alpha +1}{2}  . 
\end{equation}
For $\alpha=-\frac{1}{2}$, $J_{-\frac{1}{2}}(z)=\sqrt{\frac{2}{{\pi}z}}cos(z)$, 
hence $j_{\alpha,m} =\pi(m-\frac{1}{2})$, $m=1,2\dots,\frac{n}{2}$ and 
$k_{\frac{n}{2},\alpha}=\frac{n}{2}+\frac{1}{4}$. Therefore
\begin{equation}
(z_{m}^{n})^{2}= t_{m,\alpha}^{\frac{n}{2}}= \frac{{{\pi}^{2}}(m-\frac{1}{2})^{2}}{2n+1} \left( 1+\frac{{{\pi}^{2}}(m-\frac{1}{2})^{2} -
\frac{3}{2}}{3(2n+1)^{2}} \right)+O(n^{-5}),
\end{equation}
and the positive roots are
\begin{equation} 
z_{m}^{n}=\frac{\pi(m-\frac{1}{2})}{\sqrt{2n+1}} \sqrt{ 1+\frac{{{\pi}^{2}}(m-\frac{1}{2})^{2} -\frac{3}{2}}{3(2n+1)^{2}} }
+O(n^{-4.5}).  
\end{equation}
For an odd order, $n$, analogous relations read
\begin{eqnarray} 
H_{n} (z) &=& (-1)^{ \frac{n-1}{2} } 2^{n} { (\frac{n-1}{2}) }!z L_{{ \frac{n-1}{2} }}^{\frac{1}{2}} (z^{2}) ,
 \\
t_{m,\alpha}^{\frac{n-1}{2}}&=& \frac{{j_{m,\alpha}}^{2}}{4k_{\frac{n-1}{2},\alpha}} 
\left( 1+\frac{2({\alpha}^{2} - 1)+{j_{m,\alpha}}^{2}}{ 48{k_{\frac{n-1}{2},\alpha}}^{2}} \right)+O(n^{-5}), 
 \\
 k_{\frac{n-1}{2},\alpha}&=&\frac{n-1}{2}+\frac{\alpha +1}{2},\;\;\;\; 
 (z_{m,\alpha}^{n})^{2} =  t_{m,\alpha}^{\frac{n-1}{2}} .
   \end{eqnarray} 
\noindent  In this case $ J_{\alpha} (z)=J_{\frac{1}{2}} (z)=\sqrt{\frac{2}{{\pi}z}}sin(z)$, therefore 
$j_{\alpha,m}=\pi m$, $m=0,1,2,\dots, \frac{n-1}{2}$, and 
$k_{\frac{n-1}{2},\alpha}=\frac{n}{2}+\frac{1}{4} $. 
Then, similar steps as above give
\begin{equation} 
z_{m}^{n}=\frac{\pi m}{\sqrt{2n+1}} \sqrt{ 1+\frac{{{\pi}^{2}}m^{2} -\frac{3}{2}}{3(2n+1)^{2}} }
+O(n^{-4.5}).  
\end{equation}

\end{document}